\documentclass[12pt,reqno]{amsart}

\usepackage{fullpage}
\usepackage{graphicx}
\usepackage{amssymb, amsmath, amsxtra}

\newtheorem{thm}{Theorem}[section]
\theoremstyle{plain}

\theoremstyle{plain} 
\newtheorem{lem}{Lemma}[section]
\theoremstyle{plain}

\theoremstyle{plain}

\newtheorem{prop}{Proposition}[section]

\def\a{a}
\def\b{b}
\def\c{f}
\def\f{p}
\def\d{g}
\def\g{q}

\def\E{{\mathcal{E}}}
\def\X{{\rm X}}

\def\C{{\mathcal {C}}}

\def\parder#1{\partial_{#1}}

\def\const{\text{const.}}
\def\ie/{i.e.}
\def\eg/{e.g.}

\def\Ref#1{Ref.\cite{#1}}

\begin{document}
\allowdisplaybreaks[3]

\title{
Conservation laws of coupled semilinear\\
wave equations
}

\author{
Stephen C. Anco$^1$, 
Chaudry Masood Khalique$^{2}$\\
\\\lowercase{\scshape{
${}^1$
Department of Mathematics and Statistics, 
Brock University\\
St. Catharines, ON L2S3A1, Canada}} \\
\lowercase{\scshape{
${}^2$
International Institute for Symmetry Analysis and Mathematical Modelling\\
Department of Mathematical Sciences\\
North-West University\\
Mafikeng Campus, South Africa}}
%emails: sanco@brocku.ca, Masood.Khalique@nwu.ac.za
}

\begin{abstract}
A complete classification of all low-order conservation laws is carried out
for a system of coupled semilinear wave equations 
which is a natural two-component generalization of 
the nonlinear Klein-Gordon equation. 
The conserved quantities defined by these conservation laws 
are derived and their physical meaning is discussed. 
\end{abstract}

\maketitle

\section{Introduction}

We study a system of coupled semilinear wave equations
\begin{subequations}\label{uveqns}
\begin{align}
& u_{tt} - \a^2 u_{xx} + \c(u) +\f(v)=0,
\label{ueqn}\\
& v_{tt} - \b^2 v_{xx} + \d(v) +\g(u)=0,
\label{veqn}
\end{align}
\end{subequations}
where $a>0$ and $b>0$ are constant wave speeds, 
$\c(u)$ and $\d(v)$ are self-interaction terms, 
$\f(v)$ and $\g(u)$ are nonlinear coupling terms, 
such that $\f''(v)\neq 0$, $\g''(u)\neq 0$, and $\c(0)=\f(0)=\d(0)=\g(0)=0$. 
With these conditions, the system is 
nonlinear (\ie/ both equations contain a nonlinear term), 
coupled (\ie/ neither equation is decoupled),
and homogeneous (\ie/ $u=v=0$ is a solution). 

In general this nonlinear system 
does not have a Lagrangian formulation. 
When $\a=\b$ and $\c'=\d'=\const$, however, 
there is a Lagrangian given by 
\begin{equation}\label{L}
L = - u_t v_t + c^2 u_xv_x \pm m^2 uv +\int \f(v)\, dv +\int \g(u)\, du
\end{equation}
with $c=\const$ and $m=\const$, 
which yields 
\begin{equation}\label{uvLsys}
0=\frac{\delta L}{\delta u} = v_{tt} - c^2 v_{xx} \pm m^2 v +\g(u),
\quad
0=\frac{\delta L}{\delta v} = u_{tt} - c^2 u_{xx} \pm m^2 u +\f(v). 
\end{equation}
Note that these Euler-Lagrange equations are twisted in the sense that 
the variational derivative of $L$ with respect to $u,v$ yields 
the wave equations for $v,u$, respectively. 
In both equations, 
$c>0$ is the constant wave speed and $m\geq0$ is the mass coefficient. 

The semilinear system \eqref{uveqns} is interesting 
as it is a natural two-component generalization of 
the nonlinear Klein-Gordon equation,
where each wave component has a different speed. 
Systems of this type arise in many physical applications 
(see \eg/ \Ref{KhuPel}),
and analytical aspects of these systems have been studied recently \cite{KhuPel,Ger}. 
Conservation laws have been found in the case of cubic and other power nonlinearities \cite{BisKar}. 

The aim in the present paper is to derive a complete classification of 
all low-order conservation laws 
for the semilinear system \eqref{uveqns}
without special assumptions on the nonlinearities. 
Conservation laws are important for many reasons:
they yield conserved quantities and constants of motion; 
they can detect integrability; 
they provide potentials and nonlocally-related systems;
they can be used to check accuracy of numerical solution methods;
and they provide a way to develop good discretizations. 

In section~\ref{prelim}, 
we apply the direct method of multipliers 
\cite{AncBlu97,AncBlu02a,AncBlu02b,2ndbook} 
to set up the standard determining equations for finding the conservation laws
admitted by the semilinear system \eqref{uveqns}. 

In section~\ref{results},
we first present the classification of all low-order conservation laws and their multipliers. 
Then we examine the conserved quantities defined by these conservation laws,
and we discuss their physical meaning 
as well as their connection to variational symmetries in the case when 
the semilinear system \eqref{uveqns} has a Lagrangian formulation \eqref{uvLsys}.

Several interesting results are obtained. 
In the case when the wave speeds are equal, $\a=\b$, 
we find that the semilinear system \eqref{uveqns} 
has conserved quantities
representing energy, momentum, and boost momentum. 
This case also has a conserved energy-momentum quantity that depends on 
an arbitrary function of light-cone derivatives of $u,v$ 
when the nonlinearities are related by 
$\c(u) =(1/\alpha)\g(u)$, $\d(v) =\alpha\f(v)$, with $\alpha=\const$. 
In the Lagrangian case, 
we find that for nonlinearities given by powers of $u,v$, 
the semilinear system \eqref{uvLsys} 
has two additional conserved quantities 
representing a dilational energy-momentum and a $SO(2)$ rotation charge. 
Somewhat surprisingly, 
without any conditions relating the wave speeds $\a$ and $\b$, 
the semilinear system \eqref{uveqns} also admits 
elementary conserved quantities, which are linear in derivatives of $u,v$. 

In section~\ref{remarks}, 
we make some concluding remarks.

\section{Conservation laws}
\label{prelim}

A {\em conservation law} of the semilinear system \eqref{uveqns}
is a space-time divergence such that
\begin{equation}\label{conslaw}
D_t T(t,x,u,v,u_t,v_t,u_x,v_x,\ldots) +D_x X(t,x,u,v,u_t,v_t,u_x,v_x,\ldots)
=0
\end{equation}
holds for all solutions $(u(t,x),v(t,x))$ of the system \eqref{uveqns}. 
The spatial integral of the conserved density $T$ formally satisfies
\begin{equation}
\frac d{dt} \int_{-\infty}^{\infty} T dx 
= -X\Big|_{-\infty}^{\infty}
\end{equation}
and so if the spatial flux $X$ vanishes at spatial infinity, 
then 
\begin{equation}\label{C}
\C[u,v]= \int_{-\infty}^{\infty} T dx=\const
\end{equation}
formally yields a conserved quantity 
for the semilinear system \eqref{uveqns}.
Conversely, any such conserved quantity arises from
a conservation law \eqref{conslaw}.
Two conservation laws are equivalent if 
their conserved densities $T(t,x,u,v,u_t,v_t,u_x,v_x,\ldots)$
differ by a total $x$-derivative $D_x\Theta(t,x,u,v,u_t,v_t,u_x,v_x,\ldots)$
on all solutions $(u(t,x),v(t,x))$,
thereby giving the same conserved quantity $\mathcal C[u,v]$ up to boundary terms.
Correspondingly, 
the fluxes $X(t,x,u,v,u_t,v_t,u_x,v_x,\ldots)$ of two equivalent conservation laws
differ by a total time derivative $-D_t\Theta(t,x,u,v,u_t,v_t,u_x,v_x,\ldots)$ 
on all solutions $(u(t,x),v(t,x))$.
A conservation law is called {\em locally trivial} if 
\begin{equation}\label{trivconslaw}
T = \Phi + D_x\Theta,
\quad
X = \Psi - D_t\Theta
\end{equation}
such that $\Phi=\Psi=0$ holds on all solutions $(u(t,x),v(t,x))$.
Thus, equivalent conservation laws differ by a locally trivial conservation law.

The set of all conservation laws (up to equivalence) 
admitted by the semilinear system \eqref{uveqns} 
forms a vector space on which there is a natural action 
\cite{Olv,KarMah,BluTemAnc}
by the group of all Lie symmetries of the system \eqref{uveqns}. 

Each conservation law \eqref{conslaw} has 
an equivalent {\em characteristic form} 
in which $u_{tt}$, $v_{tt}$, and all derivatives of $u_{tt}$ and $v_{tt}$ 
are eliminated from $T$ and $X$ through use of the system \eqref{uveqns} 
and its differential consequences. 
There are two steps to obtaining the characteristic form. 
Let $\E$ denote the space of all solutions $(u(t,x),v(t,x))$ of the system \eqref{uveqns}. 
First, 
we eliminate $u_{tt},v_{tt},u_{ttx},v_{ttx},\ldots$ to get 
\begin{equation}
\hat T = T\big|_\E
= T - \Phi,
\quad
\hat X = X\big|_\E
= X - \Psi
\end{equation}
where $\hat T$ and $\hat X$ are functions only of 
$t,x,u,v,u_t,v_t,u_x,v_x,u_{tx},v_{tx},u_{xx},v_{xx},u_{txx},v_{txx},\ldots$, 
so that, on all solutions of the semilinear system \eqref{uveqns},  
\begin{equation}\label{charconslaw}
\big( D_t\hat T + D_x\hat X \big)\big|_\E =0
\end{equation}
holds with 
\begin{align}
& \begin{aligned}
D_t\big|_\E & = 
\parder{t} + u_t\parder{u}+ v_t\parder{v} 
+ u_{tx}\parder{u_x}+ v_{tx}\parder{v_x} +\cdots
+ \big(\a^2 u_{xx} -\c(u) -\f(v)\big)\parder{u_t}
\\&\qquad
+ \big(\b^2 v_{xx} -\d(v) -\g(u)\big)\parder{v_t}
+ D_x\big(\a^2 u_{xx} -\c(u) -\f(v)\big)\parder{u_{tx}} 
\\&\qquad
+ D_x\big(\b^2 v_{xx} -\d(v) -\g(u)\big)\parder{v_{tx}} 
+\cdots
\end{aligned}
\\
& D_x\big|_\E =D_x . 
\end{align}
Next, moving off of solutions, 
we use the identity 
\begin{equation}
\begin{aligned}
D_t = & D_t\big|_\E
+ \big(u_{tt} - \a^2 u_{xx} + \c(u) +\f(v)\big)\parder{u_t}
+ \big(v_{tt} - \b^2 v_{xx} + \d(v) +\g(u)\big)\parder{v_t}
\\&\qquad
+ D_x\big(u_{tt} - \a^2 u_{xx} + \c(u) +\f(v)\big)\parder{u_{tx}}
+ D_x\big(v_{tt} - \b^2 v_{xx} + \d(v) +\g(u)\big)\parder{v_{tx}}
\\&\qquad
+\cdots . 
\end{aligned}
\end{equation}
This yields the characteristic form of the conservation law \eqref{conslaw}
\begin{equation}\label{chareqn}
D_t\hat T +D_x\big(\hat X +\hat \Psi\big)
=\big(u_{tt} - \a^2 u_{xx} + \c(u) +\f(v)\big)Q^u + \big(v_{tt} - \b^2 v_{xx} + \d(v) +\g(u)\big)Q^v
\end{equation}
holding identically, 
where
\begin{equation}\label{trivX}
\begin{aligned}
\hat\Psi & = 
E_{u_{tx}}(\hat T)\big(u_{tt} - \a^2 u_{xx} + \c(u) +\f(v)\big)
+E_{v_{tx}}(\hat T)\big(v_{tt} - \b^2 v_{xx} + \d(v) +\g(u)\big)
\\&\qquad
+ E_{u_{txx}}(\hat T)D_x(u_{tt} - \a^2 u_{xx} + \c(u) +\f(v)\big)
+E_{v_{txx}}(\hat T)D_x\big(v_{tt} - \b^2 v_{xx} + \d(v) +\g(u)\big)
\\&\qquad
+\cdots
\end{aligned}
\end{equation}
is a trivial flux, 
and where the pair of functions
\begin{equation}\label{TQrelation}
Q^u=E_{u_t}(\hat T), 
\quad
Q^v=E_{v_t}(\hat T) 
\end{equation}
is called a {\em multiplier} (or a {\em characteristic}). 
Here $E_w=\parder{w}-D_x\parder{w_x} +D_x^2\parder{w_{xx}} -\cdots$
denotes the (spatial) Euler operator \cite{Olv,2ndbook} 
with respect to a variable $w$. 
This operator has the important property \cite{Olv,2ndbook}
that a function $h(t,x,w,w_x,w_{xx},\ldots)$
is annihilated by $E_w$ iff the function is a total (spatial) derivative, 
$h=D_x\theta(t,x,w,w_x,w_{xx},\ldots)$. 

The relation \eqref{TQrelation} between $\hat T$ and $(Q^u,Q^v)$ 
shows that a multiplier will be uniquely determined as a function of 
$t,x,u,v,u_t,v_t,u_x,v_x,u_{tx},v_{tx},u_{xx},v_{xx},u_{txx},v_{txx},\ldots$ 
if $\hat T$ has no dependence on $u_{tt}$, $v_{tt}$, and their derivatives. 
In particular, the multiplier for a locally trivial conserved density 
$\hat T=D_x\Theta$ vanishes, since $E_{u_t}(D_x\Theta)= E_{v_t}(D_x\Theta)= 0$. 
Conversely, for a trivial multiplier $(Q^u,Q^v)=0$, 
the relation \eqref{TQrelation} implies that $\hat T=D_x\Theta + \hat T_0$
holds for some function $\hat T_0(t,x,u,v,u_x,v_x,u_{xx},v_{xx},\ldots)$
with no dependence on $u_t$, $v_t$, and their derivatives.
Then the characteristic equation \eqref{chareqn} yields the relation
$D_t\hat T_0 + D_x\hat X_0=0$, which holds identically,
where $\hat X_0=D_t\Theta -\hat X$. 
This determines $\hat T_0=D_x\Theta_0$ and $\hat X_0=-D_t\Theta_0$, 
showing that $\hat T=D_x(\Theta +\Theta_0)$ is locally trivial. 

Thus, there is a one-to-one relation between 
multipliers and conserved densities (up to equivalence). 
All multipliers
\begin{equation}\label{genQ}
\begin{aligned}
& (Q^u(t,x,u,v,u_t,v_t,u_x,v_x,u_{tx},v_{tx},u_{xx},v_{xx},u_{txx},v_{txx},
\ldots),
\\&\qquad 
Q^v(t,x,u,v,u_t,v_t,u_x,v_x,u_{tx},v_{tx},u_{xx},v_{xx},u_{txx},v_{txx},\ldots))
\end{aligned}
\end{equation}
(with a finite differential order) 
are determined by the condition that 
their summed product with the equations in the semilinear system \eqref{uveqns}
is a total space-time divergence. 
Such divergences have the characterization that their variational derivative
with respect to $u$ and $v$ vanishes identically \cite{Olv,2ndbook}. 
This condition 
\begin{subequations}\label{Qvardercond}
\begin{align}
& \frac{\delta}{\delta u}\Big(
\big(u_{tt} - \a^2 u_{xx} + \c(u) +\f(v)\big)Q^u + \big(v_{tt} - \b^2 v_{xx} + \d(v) +\g(u)\big)Q^v
\Big)=0
\\
&
\frac{\delta}{\delta v}\Big(
\big(u_{tt} - \a^2 u_{xx} + \c(u) +\f(v)\big)Q^u + \big(v_{tt} - \b^2 v_{xx} + \d(v) +\g(u)\big)Q^v
\Big)=0
\end{align}
\end{subequations}
can be split in an explicit form with respect to $u_{tt},v_{tt},u_{ttx},v_{ttx},u_{ttxx},v_{ttxx},\ldots$,
which yields an equivalent set of equations 
\cite{AncBlu97,AncBlu02a,AncBlu02b} 
\begin{equation}\label{adjsymmdeteq}
D_t^2 Q^u - a^2D_x^2 Q^u +\c'(u)Q^u +\g'(u)Q^v=0,
\quad
D_t^2 Q^v -b^2D_x^2 Q^v +\d'(v)Q^v+\f'(v)Q^u=0, 
\end{equation}
and
\begin{equation}\label{Helmholtzeq}
\begin{aligned}
& 
\parder{u_t}Q^u = E_{u_t}(Q^u),
\quad
\parder{u_t}Q^v = E_{v_t}(Q^u),
\quad
\parder{v_t}Q^u = E_{u_t}(Q^v),
\quad
\parder{v_t}Q^v = E_{v_t}(Q^v),
\\
& 
\parder{u_{tx}}Q^u = -E_{u_t}^{(1)}(Q^u),\ 
\parder{u_{tx}}Q^v = -E_{v_t}^{(1)}(Q^u),\ 
\parder{v_{tx}}Q^u = -E_{u_t}^{(1)}(Q^v),\
\parder{v_{tx}}Q^v = -E_{v_t}^{(1)}(Q^v),
\\
&
\parder{u_{txx}}Q^u = -E_{u_t}^{(2)}(Q^u),\ 
\parder{u_{txx}}Q^v = -E_{v_t}^{(2)}(Q^u),\ 
\parder{v_{txx}}Q^u = -E_{u_t}^{(2)}(Q^v),\ 
\parder{v_{txx}}Q^v = -E_{v_t}^{(2)}(Q^v),
\\
&
\text{ etc. }
\end{aligned}
\end{equation}
holding for all solutions $(u(t,x),v(t,x))$ 
of the semilinear system \eqref{uveqns}.
Here $E_w^{(1)} = \parder{w_x}-2D_x\parder{w_{xx}} +3D_x^2\parder{w_{xxx}} -\cdots$, 
$E_w^{(2)} = \parder{w_{xx}}-3D_x\parder{w_{xxx}} +6D_x^2\parder{w_{xxxx}} -\cdots$,
etc.\
denote higher (spatial) Euler operators \cite{Olv}. 
These equations \eqref{adjsymmdeteq}--\eqref{Helmholtzeq} constitute
the standard determining system for multipliers \cite{AncBlu02a,AncBlu02b,2ndbook}. 

In the case when the semilinear system \eqref{uveqns} 
is a Lagrangian system \eqref{uvLsys},
the first equation \eqref{adjsymmdeteq} 
is simply the determining equation for symmetries 
$\X=Q^u\parder{u}+Q^v\parder{v}$ in evolutionary form
\cite{Olv,1stbook,2ndbook}. 
The second equation \eqref{Helmholtzeq} is then equivalent to the condition 
that the Lagrangian \eqref{L} is invariant under $\X=Q^u\parder{u}+Q^v\parder{v}$ 
to within a total space-time divergence. 

Hence, in the Lagrangian case, 
the determining equations \eqref{adjsymmdeteq} and \eqref{Helmholtzeq} 
for multipliers are equivalent to the condition \cite{BluCheAnc}
that $\X=Q^u\parder{u}+Q^v\parder{v}$ is a variational symmetry 
of the semilinear system. 

In the general case, 
the first equation \eqref{adjsymmdeteq} instead is the adjoint of the symmetry determining equation, 
and its solutions $(Q^u,Q^v)$ are called {\em adjoint-symmetries}. 
The second equation \eqref{Helmholtzeq} then comprises the Helmholtz conditions
which are necessary and sufficient for $(Q^u,Q^v)$ to be 
an Euler-Lagrange expression \eqref{TQrelation}. 
Consequently, in this case \cite{Zha,Olv,AncBlu02a,AncBlu02b}, 
multipliers are simply adjoint-symmetries that have a variational form,
and the determination of conservation laws via multipliers is
a kind of adjoint problem \cite{AncBlu97} of the determination of symmetries. 

For any solution of the multiplier determining equations \eqref{adjsymmdeteq} and \eqref{Helmholtzeq}, 
a conserved density and a flux can be recovered 
either by \cite{Wol02b,2ndbook} 
directly integrating the relation \eqref{TQrelation} between $(Q^u,Q^v)$ and $\hat T$, 
or by \cite{Olv,AncBlu02a,AncBlu02b} 
using a homotopy integral formula which expresses $\hat T$ in terms of $(Q^u,Q^v)$
(see also \Ref{Anc,PooHer}). 

A conservation law of the semilinear system \eqref{uveqns} 
is said to be of {\em low order} \cite{Anc-capetown}
if the only derivatives of $u,v$ that appear in its multiplier 
are related to the leading derivatives $u_{tt},v_{tt},u_{xx},v_{xx}$ in the system 
by differentiation with respect to $t,x$. 
This means that the multiplier for a low-order conservation law is at most first-order, 
\begin{equation}\label{loworderQ}
Q=(Q^u(t,x,u,v,u_t,v_t,u_x,v_x),Q^v(t,x,u,v,u_t,v_t,u_x,v_x)) . 
\end{equation}
In general, for wave equations, 
conservation laws of physical importance, such as energy and momentum, 
are of low order,
while conservation laws connected with integrability are typically of higher order.

\section{Classification results}
\label{results}

We will now classify all low-order conservation laws \eqref{charconslaw}
admitted by the semilinear system \eqref{uveqns}. 
This means finding all multipliers \eqref{TQrelation} 
with a general first-order form \eqref{loworderQ}. 
The corresponding conserved densities and fluxes will then have the following 
general form. 

From relation \eqref{TQrelation} 
and properties of the (spatial) Euler operator, 
a first-order multiplier \eqref{loworderQ} determines that a conserved density 
has the form $\hat T= T(t,x,u,v,u_t,v_t,u_x,v_x) + T_0 + D_x\Theta$
where, without loss of generality, 
the function $T_0$ contains no purely first-order terms
and no $t$-derivatives of $u,v$. 
The characteristic equation \eqref{chareqn} then yields 
$D_t|_\E T + D_t T_0 + D_x\hat X = 0$. 
This relation splits with respect to all derivatives of $u,v$ 
with order greater than two. 
Starting at the highest order terms, 
the splitting leads to $T_0 = D_x\Theta+\Phi$ and $\hat X= -D_t\Theta +X+\Psi$
by a standard descent argument in derivatives of $u,v$,
where $\Phi|_\E=0$ and $\Psi|_\E=0$ are trivial terms, 
and where $X(t,x,u,v,u_t,v_t,u_x,v_x)$ is first-order. 
Hence, the non-trivial part of $\hat T$ and $\hat X$ consists of 
only the first-order terms $T$ and $X$. 
Conversely, 
a first-order conserved density $\hat T(t,x,u,v,u_t,v_t,u_x,v_x)$
and a first-order flux $\hat X(t,x,u,v,u_t,v_t,u_x,v_x)$
determine a first-order multiplier \eqref{loworderQ} 
directly by the relation \eqref{TQrelation}. 
Thus, we have established the following characterization of low-order conservation laws. 

\begin{lem}
For the semilinear system \eqref{uveqns}, 
a non-trivial conservation law \eqref{charconslaw}
will have a multiplier of first-order \eqref{loworderQ} 
iff the conserved density and the flux, up to equivalence, 
are of first-order 
\begin{equation}\label{loworderTX}
\hat T=T(t,x,u,v,u_t,v_t,u_x,v_x),
\quad
\hat X=X(t,x,u,v,u_t,v_t,u_x,v_x) .
\end{equation}
Therefore, the class of low-order conservation laws consists of 
all admitted non-trivial conserved densities and fluxes 
with the form \eqref{loworderTX}.
\end{lem}

\subsection{Multipliers, conserved densities and fluxes}

The classification of low-order conservation laws 
will be carried out modulo the equivalence transformations
\begin{equation}\label{reflect}
\begin{aligned} 
& u\rightarrow v, 
\quad
\c(u)\rightarrow \g(u),
\quad
\f(v)\rightarrow \d(v),
\\
& v\rightarrow u,
\quad
\g(u)\rightarrow \c(u),
\quad
\d(v)\rightarrow \f(v)
\end{aligned}
\end{equation}
and
\begin{equation}\label{scal}
\begin{aligned} 
& u\rightarrow \alpha u, 
\quad
\c(u)\rightarrow \alpha\c(\alpha u),
\quad
\f(v)\rightarrow \alpha\f(\beta v),
\quad
\alpha\neq0
\\
& v\rightarrow \beta v,
\quad
\g(u)\rightarrow \beta\g(\alpha u),
\quad 
\d(v)\rightarrow \beta\f(\beta v),
\quad
\beta\neq0
\end{aligned}
\end{equation}
under which the general form of the semilinear system \eqref{uveqns} 
is preserved. 

Based on the form of the equations \eqref{ueqn}--\eqref{veqn} in the system, 
it is computationally convenient to divide the classification 
into three main cases:
(I) $\c(u)=\d(v)=0$; 
(II) $\c(u)\neq 0$, $\d(v)=0$; or $\c(u)=0$, $\d(v)\neq0$; 
(III) $\c(u)\neq 0$, $\d(v)\neq 0$.
Note, under a reflection transformation \eqref{reflect}, 
the case $\c(u)=0$, $\d(v)\neq0$ is equivalent to the case $\c(u)\neq 0$, $\d(v)=0$. 
For each case, 
we use Maple to set up and solve the determining system for multipliers. 

\begin{prop}\label{classI-multipliers}
For the semilinear system of wave equations \eqref{uveqns}
in the case $\c(u)=\d(v)=0$,
the admitted conservation law multipliers consist of:
\begin{align}
{\rm (a)}\qquad
& 
\a=\b, 
\quad
\text{arbitrary } \f(v),\g(u)
\nonumber\\
& 
Q =(\b^2 t v_x + x v_t, \b^2 t u_x + x u_t)
\label{QIa1}
\\
& 
Q =(v_x, u_x)
\label{QIa2}
\\
&
Q =(v_t, u_t)
\label{QIa3}
\\
{\rm (b)}\qquad
& 
\a=\b,
\quad
\f(v)= \beta v^{k-1},
\quad
\g(u)=\alpha u^{-k-1}
\nonumber\\
&
Q =(k t v_t + k x v_x+2 v, k t u_t + k x u _x -2 u)
\label{QIb1}
\\
{\rm (c)}\qquad
& 
\a=\b,
\quad
\f(v)= \beta e^{\delta v},
\quad  
\g(u)=\alpha e^{\gamma u}
\nonumber\\
& 
Q = ( (v_t\pm \b v_x)A(t,x) \pm 2(\b/\delta) A_x(t,x), (u_t\pm \b u_x)A(t,x) \pm 2(\b/\gamma) A_x(t,x) )
\label{QIc1}
\\&\text{where }
A_t\mp \b A_x =0 
\nonumber
\end{align}
\end{prop}

\begin{prop}\label{classII-multipliers}
For the semilinear system of wave equations \eqref{uveqns}
in the case $\c(u)\neq 0$ and $\d(v)=0$, 
the admitted conservation law multipliers consist of:
\begin{align}
{\rm (a)}\qquad
& 
\a=\b,
\quad
\c(u)=\kappa\g(u),
\quad
\text{arbitrary } \f(v),\g(u)
\nonumber\\
&
Q=(\b^2 t v_x + x v_t, \b^2 t(u_x - \kappa v_x) + x(u_t -\kappa v_t) )
\label{QIIa1}
\\
&
Q=(v_x, u_x -\kappa v_x)
\label{QIIa2}
\\
&
Q=(v_t, u_t -\kappa v_t)
\label{QIIa3}
\\
{\rm (b)}\qquad
& 
\a=\b,
\quad
\c(u)=\kappa \g(u),
\quad
\f(v)= \beta e^{\delta v},
\quad  
\g(u)=\alpha e^{\gamma u}
\nonumber\\
& 
\begin{aligned}
Q & = ( (v_t\pm \b v_x)A(t,x) +(2/\delta) A_t(t,x), 
\\&\qquad
(u_t-\kappa v_t \pm\b (u_x-\kappa v_x))A(t,x) +2(1/\gamma-\kappa/\delta) A_t(t,x) )
\end{aligned}
\label{QIIb1}
\\
&\text{where }
A_t\mp \b A_x =0
\nonumber
\end{align}
\end{prop}

\begin{prop}\label{classIII-multipliers}
For the semilinear system of wave equations \eqref{uveqns}
in the case $\c(u)\neq 0$ and $\d(v)\neq 0$,
the admitted conservation law multipliers consist of:
\begin{align}
{\rm (a)}\qquad
& 
\a\neq \b,
\quad
\c(u) = (1/\alpha) \g(u) + \delta u,
\quad
\d(v) = \alpha\f(v) + \gamma v,  
\quad
\text{arbitrary } \f(v),\g(u)
\nonumber\\
&
Q = (-\alpha A(t)B(x),A(t)B(x))
\label{QIIIa1}
\\
&\text{where } 
(\a^2 - \b^2)A'' + (\a^2 \gamma -\b^2 \delta)A=0,
\quad
(\a^2 - \b^2)B'' + (\gamma -\delta)B=0
\nonumber
\\
{\rm (b)}\qquad
& 
\a=\b,
\quad
\c(u) = \beta \g(u) + \gamma u,
\quad
\d(v) = \alpha \f(v) + \gamma v,
\quad
\text{arbitrary } \f(v),\g(u)
\nonumber\\
&
Q = (\b^2 t(\alpha u_x - v_x) + x(\alpha u_t - v_t),\b^2 t(\beta v_x - u_x) + x(\beta v_t - u_t))
\label{QIIIb1}
\\
& 
Q =(\alpha u_x - v_x,\beta v_x - u_x)
\label{QIIIb2}
\\
& 
Q = (\alpha u_t - v_t,\beta v_t - u_t)
\label{QIIIb3}
\\
{\rm (c)}\qquad
& 
\a=\b,
\quad
\c(u) = (1/\alpha)\g(u) + \gamma u,
\quad
\d(v) = \alpha \f(v) + \gamma v,
\quad
\text{arbitrary } \f(v),\g(u)
\nonumber\\
&
Q = (-\alpha A(t,x), A(t,x))
\label{QIIIc1}
\\&\text{where }
A_{tt} -\b^2 A_{xx} +\gamma A=0
\nonumber
\\
{\rm (d)}\qquad
& 
\a=\b,
\quad
 \c(u) =(1/\alpha)\g(u),
\quad
\d(v) =\alpha\f(v), 
\quad
\text{arbitrary } \f(v),\g(u)
\nonumber\\
&
Q = ( -\alpha A(\zeta), A(\zeta))
\label{QIIId1}
\\&\text{where }
\zeta=\alpha u_t-v_t \pm\b(\alpha u_x-v_x)
\nonumber
\\
{\rm (e)}\qquad
& 
\a=\b,
\quad
\c(u) =  \beta \g(u), 
\quad
\d(v) = \alpha \f(v),
\quad
\f(v) = \kappa e^ {\mu v},
\quad
\g(u) = \lambda e^{\delta u} 
\nonumber\\
&
\begin{aligned}
Q & = ( (\alpha u_t - v_t\pm\b(\alpha u_x - v_x)) A(t,x) +2(\alpha/\delta-1/\mu)  A_t(t,x),
\\&\qquad
(\beta v_t- u_t \pm\b(\beta v_x -u_x)) A(t,x) + 2(\beta/\mu -1/\delta) A_t(t,x) )
\end{aligned}
\label{QIIIe1}
\\&\text{where }
A_t\mp \b A_x =0
\nonumber
\\
{\rm (f)}\qquad
& 
\a=\b,
\quad
\c(u) =  \gamma u,
\quad
\d(v) = \gamma v,
\quad
\f(v) = \alpha/v,
\quad
\g(u) = \beta/u
\nonumber\\
&
Q = (-v,u)
\label{QIIIf1}
\end{align}
\end{prop}

Each multiplier $Q$ determines 
a conserved density $\hat T$ and a flux $\hat X$, 
up to equivalence. 
The simplest way to obtain explicit expressions for them is by 
first splitting the characteristic equation \eqref{chareqn}
with respect to $u_{xx}$, $v_{xx}$, $u_{tx}$, $v_{tx}$, 
and next integrating the resulting linear system 
\begin{gather}
\hat T_{u_t} - Q^u =0,
\quad
\hat T_{v_t} - Q^v =0,
\\
\hat X_{u_t}+ \hat T_{u_x}=0,
\quad
\hat X_{v_t} + \hat T_{v_x} =0,
\\
\hat X_{u_x}+ \a^2\hat T_{u_t}=0,
\quad
\hat X_{v_x} + \b^2\hat T_{v_t} =0,
\\
\hat T_{t} + u_t \hat T_{u} + v_t \hat T_{v} 
+ \hat X_{x} + u_x \hat X_{u} + v_x \hat X_{v} 
-(\c(u) +\f(v))\hat T_{u_t}-(\d(v) +\g(u))\hat T_{v_t}
=0 . 
\end{gather}
This leads to the following main classification result, 
after the various overlapping cases in 
Propositions~\ref{classI-multipliers}, ~\ref{classII-multipliers}, and~\ref{classIII-multipliers}
have been merged. 

\begin{thm}\label{class-conslaws}
All low-order conservation laws \eqref{loworderTX} admitted by 
the semilinear system of wave equations \eqref{uveqns}
are given by, up to equivalence, the conserved densities and the fluxes:
\begin{align}
{\rm (a)}\qquad
& 
\a\neq \b,
\quad
\c(u) = (1/\alpha)\g(u) + \delta u,
\quad
\d(v) = \alpha \f(v) + \gamma v,  
\quad
\text{arbitrary } \f(v),\g(u)
\nonumber\\
&
\begin{aligned} 
T_{1} & = 
\big( (v_t -\alpha u_t) A(t) +(\alpha u-v)A'(t) \big) B(x)
\\
X_{1} & = 
\big( (\alpha\a^2 u_x -\b^2 v_x) B(x) + (\b^2 v-\alpha\a^2 u) B'(x) \big) A(t)
\end{aligned}
\label{TXa1}
\\
&\qquad\text{ where }
(\a^2 - \b^2)A'' + (\a^2 \gamma -\b^2 \delta)A=0,
\quad
(\a^2 - \b^2)B'' + (\gamma -\delta)B=0
\nonumber
\\
{\rm (b)}\qquad
& 
\a=\b,
\quad
\c(u) = \beta \g(u) + \gamma u,
\quad
\d(v) = \alpha \f(v) + \gamma v,
\quad
\text{arbitrary } \f(v),\g(u)
\nonumber\\
&
\begin{aligned} 
T_{2} &= 
(\alpha\beta - 1) \left( \int f(v) dv + \int g(u) du \right) 
- (u_t v_t +\b^2 u_x v_x + \gamma u v)
\\&\qquad
+ \tfrac{1}{2} \alpha (v_t^2 +\b^2 v_x^2 + \gamma v^2)
+ \tfrac{1}{2} \beta (u_t^2+\b^2 u_x^2 + \gamma u^2) 
\\
X_{2} &=  
\b^2 \big( u_t(v_x-\beta u_x)  + v_t(u_x-\alpha v_x)  \big)
\end{aligned}  
\label{TXb1}
\\
&
\begin{aligned} 
T_{3} & = 
(\beta u_x - v_x)u_t + (\alpha v_x - u_x)v_t   
\\
X_{3} &= 
(\alpha\beta - 1) \left( \int f(v) dv + \int g(u) du \right) 
+ u_t v_t + \b^2 u_x v_x - \gamma u v
\\&\qquad
-\tfrac{1}{2} \alpha (v_t^2 +\b^2 v_x^2 -\gamma v^2 )
-\tfrac{1}{2} \beta (u_t^2+\b^2 u_x^2 -\gamma u^2) 
\end{aligned}
\label{TXb2}
\\
&
\begin{aligned} 
T_{4} &= 
(\alpha\beta - 1) x\left( \int f(v) dv + \int g(u) du \right) 
-x(u_t v_t +b^2 u_x v_x +\gamma u v) 
\\&\qquad
+ \tfrac{1}{2} \alpha x (u_t^2 +\b^2 u_x^2 +\gamma u^2) 
+ \tfrac{1}{2} \beta x (v_t^2 +\b^2 v_x^2 +\gamma v^2) 
\\&\qquad
+\b^2 t (\alpha u_t u_x +\beta v_t v_x - u_t v_x -u_x v_t)
\\
X_{4} &= 
\b^2 (\alpha\beta - 1) t \left( \int f(v) dv + \int g(u) du \right) 
+ \b^2 t( u_t v_t + \b^2 u_x v_x - \gamma u v) 
\\&\qquad
-\tfrac{1}{2} \alpha \b^2  t(u_t^2 +\b^2 u_x^2 - \gamma u^2)
- \tfrac{1}{2} \beta \b^2 t (v_t^2 +\b^2 v_x^2 - \gamma v^2)
\\&\qquad
+ \b^2 x (u_t v_x + u_x v_t -\alpha u_t u_x - \beta v_t v_x)
\end{aligned}
\label{TXb3}
\\
{\rm (c)}\qquad
& 
\a=\b,
\quad
\c(u) = (1/\alpha)\g(u) + \gamma u,
\quad
\d(v) = \alpha \f(v) + \gamma v,
\quad
\text{arbitrary } \f(v),\g(u)
\nonumber\\
&
\begin{aligned} 
T_{5} &= 
(\alpha u_t - v_t)A(t,x) + (v -\alpha u)A_t(t,x)
\\
X_{5} & = 
\b^2\big( (v_x-\alpha u_x)A(t,x) + (\alpha u-v)A_x(t,x) \big)
\end{aligned} 
\label{TXc1}
\\
&\qquad \text{ where } 
A_{tt}-\b^2 A_{xx} +\gamma A=0
\nonumber
\\ 
{\rm (d)}\qquad
& 
\a=\b,
\quad
 \c(u) =(1/\alpha)\g(u),
\quad
\d(v) =\alpha\f(v), 
\quad
\text{arbitrary } \f(v),\g(u)
\nonumber\\
&
\begin{aligned}
T_{6} & = 
B(\zeta)
\\
X_{6} &=  
\mp\b B(\zeta)
\end{aligned}
\label{TXd1}
\\
&\qquad \text{ where } 
\zeta= \alpha u_t-v_t \pm\b(\alpha u_x-v_x) 
\nonumber
\\
{\rm (e)}\qquad
& 
\a=\b,
\quad
\c(u) =  \beta \g(u), 
\quad
\d(v) = \alpha \f(v),
\quad
\f(v) = \kappa e^ {\mu v},
\quad
\g(u) = \lambda e^{\delta u} 
\nonumber\\
&
\begin{aligned} 
T_{7} & = 
\big( (\alpha\beta - 1)( \lambda\gamma\exp(\delta u) + \delta\kappa\exp(\gamma v) )
+\tfrac{1}{2}\delta \gamma \big( \alpha (u_t\pm\b u_x)^2 +\beta (v_t \pm\b v_x)^2 \big) 
\\&\qquad
-\delta \gamma (u_t\pm\b u_x)(v_t \pm\b v_x) 
\big) A(t,x) 
+2\big( (\delta\beta -\gamma) v_t +(\gamma\alpha -\delta) u_t \big) A_t(t,x)
\\&\qquad
+2\big((\delta -\gamma\alpha) u + (\gamma -\delta\beta) v \big) A_{tt}(t,x) 
\\
X_{7} & = 
\pm\b\Big(
\big( (\alpha\beta-1)(\lambda\gamma \exp(\delta u) +\delta\kappa \exp(\gamma v) ) 
-\tfrac{1}{2}\delta\gamma\big( \alpha(u_t\pm\b u_x)^2 +\beta(v_t \pm\b v_x)^2 \big)
\\&\qquad
+\delta\gamma (u_t\pm\b u_x)(v_t \pm\b v_x) 
\big) A(t,x) 
+2\big( (\delta - \gamma\alpha) u_x +(\gamma -\delta\beta) v_x \big) A_x(t,x) 
\\&\qquad 
+2\b^2\big( (\delta \beta -\gamma) v +(\gamma \alpha -\delta) u \big) A_{xx} (t,x) 
\Big)
\end{aligned}  
\label{TXe1}
\\
&\qquad \text{ where } 
A_t\mp \b A_x =0
\nonumber
\\ 
{\rm (f)}\qquad
& 
\a=\b,
\quad
\c(u) =  \gamma u,
\quad
\d(v) = \gamma v,
\quad
\f(v) = \alpha/v,
\quad
\g(u) = \beta/u,
\quad
\gamma\neq 0
\nonumber\\
&
\begin{aligned} 
T_{8} & =  
u v_t - v u_t 
\\
X_{8} & =  
\b^2 (v u_x - u v_x) + (\beta - \alpha)x
\end{aligned}  
\label{TXf1}
\\ 
{\rm (g)}\qquad
& 
\a=\b,
\quad
\c=\d=0,
\quad
\f(v)= \alpha v^{k-1},
\quad
\g(u)=\beta u^{-k-1}
\nonumber\\
&
\begin{aligned} 
T_{9} & = 
k t( v_t u_t +\b^2 v_x u_x ) + k x (v_x u_t + u_x v_t)
+2(v u_t-u v_t)  +t(\alpha v^{k} -\beta u^{-k})
\\
X_{9} & = 
-b^2 k t (v_t u_x +u_t v_x)  -k x (v_t u_t  +\b^2 v_x u_x) 
+x( \alpha v^{k} -\beta u^{-k} )
\end{aligned} 
\label{TXg1}
\end{align}
\end{thm}

\subsection{Conserved quantities}

We now discuss the conserved quantities \eqref{C} that arise from 
the conservation laws in Theorem~\ref{class-conslaws}
for the semilinear system \eqref{uveqns}. 

The case of unequal wave speeds $\a\neq\b$ will be considered first. 

The conservation law \eqref{TXa1} in part (a)
is a linear expression in $u,v,u_t,u_x,v_t,v_x$. 
This conservation law is admitted 
because the wave equations \eqref{ueqn} and \eqref{veqn} in the case 
$\c(u) = (1/\alpha)\g(u) + \delta u$, $\d(v) = \alpha\f(v) + \gamma v$, $\a\neq \b$
can be combined to yield a linear equation 
\begin{equation}\label{linuveqn}
(\alpha u -v)_{tt} - (\alpha\a^2 u -\b^2 v)_{xx} +\delta\alpha u - \gamma v=0 . 
\end{equation}
Any linear PDE admits elementary conserved quantities which are given by 
the projection of solutions of the PDE 
onto modes that satisfy the adjoint linear system. 
The modes for the linear equation \eqref{linuveqn} consist of 
the multiplier expression \eqref{QIIIa1} 
which has the form $Q=(-\alpha F,F)$ 
where $F(t,x)=A(t)B(x)$ is a mode function 
satisfying the adjoint of equation \eqref{linuveqn}, 
\begin{equation}
(\a^2-\b^2)F_{xx}+(\gamma-\delta)F=0,
\quad
(\a^2-\b^2)F_{tt}+(\gamma\a^2-\delta\b^2)F=0 . 
\end{equation}
From this adjoint system, 
in the general case when $\gamma\neq\delta$ and $\a^2\gamma\neq\b^2\delta$, 
the mode function is a harmonic travelling wave $F=\exp(i(k x+w t))$ 
where 
\begin{equation}
k = \sqrt{(\gamma - \delta)/(\a^2 - \b^2)}
\end{equation}
is the reciprocal wave length, 
and 
\begin{equation}
w=\sqrt{(\a^2 \gamma -\b^2 \delta)/(\a^2 - \b^2)}
\end{equation}
is the frequency, 
given in terms of the wave speeds $\a,\b$ and the mass coefficients $\delta,\gamma$ 
in the wave equations \eqref{ueqn} and \eqref{veqn}. 
The resulting conserved quantities have the form 
\begin{equation}\label{elem}
\C[u,v]= 
\int_{-\infty}^{\infty} ((v_t -\alpha u_t) +iw (\alpha u-v))\exp(i(k x+w t)) \;dx . 
\end{equation}
(Note in some cases $k$ or $w$ will be imaginary, 
so the mode function becomes an exponential in $x$ or $t$;
in all other cases, 
the mode function degenerates into a form
where its dependence on $x$ or $t$ is a linear polynomial.)

These conserved quantities \eqref{elem} are the only ones admitted by 
the semilinear system \eqref{uveqns} 
when the wave speeds are unequal, $\a\neq\b$. 
Next, the case of equal wave speeds, but with no Lagrangian being admitted,
will be considered. 

The conservation law \eqref{TXc1} in part (c) 
is a linear expression in $u,v,u_t,u_x,v_t,v_x$. 
It is admitted because the wave equations \eqref{ueqn} and \eqref{veqn} 
in the case 
$\c(u) = (1/\alpha)\g(u) + \gamma u$, $\d(v) = \alpha\f(v) + \gamma v$, $\a=\b$
can be combined into a linear wave equation for $\alpha u-v$, 
\begin{equation}\label{linwaveeqn}
(\alpha u -v)_{tt} - \b^2(\alpha u - v)_{xx} +\gamma(\alpha u -v)=0 . 
\end{equation}
Similarly to part (a), there are elementary conserved quantities 
given by the projection of solutions of this linear equation 
onto modes that satisfy the adjoint linear system. 
Since the equation \eqref{linwaveeqn} is self-adjoint, 
the modes have the form given by the multiplier expression \eqref{QIIIc1} 
where the mode function $A(t,x)$ is the general solution of the wave equation 
\begin{equation}
A_{tt} -\b^2 A_{xx}+\gamma A=0 . 
\end{equation}
Note that $\b>0$ is wave speed and $\gamma$ is the squared-mass coefficient. 
Solutions of this wave equation can be expressed as Fourier modes 
$A=\exp(i(k x \pm w t))$ where $w^2=\b^2k^2+\gamma$ 
is the standard dispersion relation, with $k$ being arbitrary. 
In the case $\gamma=0$, 
the general solution is simply a linear combination of 
left-moving and right-moving travelling waves, $A=A_\pm(x\pm \b t)$. 
In the case $\gamma\neq0$, 
the general solution depends on two arbitrary functions 
which can be identified with the amplitudes for the two Fourier modes, 
giving $A=\int_{-\infty}^\infty \exp(i(k x \pm w t)) \hat A_\pm(k)dk$. 
In both cases, 
the resulting conserved quantities are given by 
the expression \eqref{elem}, where $k$ is now an arbitrary parameter. 

The three conservation laws \eqref{TXb1}, \eqref{TXb2}, \eqref{TXb3} in part (b)
depend quadratically on $u_t,u_x,v_t,v_x$
and the first two do not contain $t,x$. 
They respectively yield a conserved energy and momentum
\begin{align}
\label{ener}
&\begin{aligned}
\C[u,v]_{\text{ener.}} & = 
\int_{-\infty}^{\infty}\bigg(  
(\alpha\beta - 1) \Big( \int f(v) dv + \int g(u) du \Big) 
+ \tfrac{1}{2} (\alpha-1) (v_t^2 +\b^2 v_x^2 + \gamma v^2)
\\&\qquad
+ \tfrac{1}{2} (\beta-1) (u_t^2+\b^2 u_x^2 + \gamma u^2) 
+ \tfrac{1}{2}\big( (u_t-v_t)^2 +\b^2(u_x-v_x)^2 + \gamma(u-v)^2 \big)
\bigg)\; dx , 
\end{aligned}
\\
\label{mom}
&\begin{aligned}
\C[u,v]_{\text{mom.}} & = 
\int_{-\infty}^{\infty}\Big(
(\alpha-1) v_t v_x + (\beta-1) u_t u_x +(u_t -v_t)(u_x -v_x)
\Big)\; dx , 
\end{aligned}
\end{align}
and a conserved boost-momentum
\begin{equation}\label{boostmom}
\begin{aligned}
\C[u,v]_{\text{boost mom.}} & = 
\int_{-\infty}^{\infty}\bigg(
(\alpha\beta - 1) x\left( \int f(v) dv + \int g(u) du \right) 
+ \tfrac{1}{2}x\big( (\alpha-1)(v_t^2 +\b^2 v_x^2 + \gamma v^2)
\\&\qquad
+ (\beta-1) (u_t^2+\b^2 u_x^2 + \gamma u^2) 
+ (u_t-v_t)^2 +\b^2(u_x-v_x)^2 + \gamma(u-v)^2 \big)
\\&\qquad
+\b^2 t\big( (\alpha-1) v_t v_x + (\beta-1) u_t u_x +(u_t -v_t)(v_x -u_x) \big)
\bigg)\; dx . 
\end{aligned}
\end{equation}
The conserved energy will be positive 
when $\alpha>1$, $\beta>1$, and $\gamma>0$. 
In contrast, 
the sign of the conserved momentum depends on the relative signs of $u_t,u_x$  
as well as $v_t,v_x$. 
When these signs are positive, corresponding to right-moving wave motion, 
the momentum is positive if $\alpha>1$ and $\beta>1$,
whereas when the signs are negative, corresponding to left-moving wave motion, 
the momentum is negative if $\alpha>1$ and $\beta>1$. 

The conservation law \eqref{TXe1} in part (e) 
also depends quadratically on $u_t,u_x,v_t,v_x$. 
It yields a conserved hyperbolic-energy 
\begin{equation}\label{hyperener}
\begin{aligned}
\C[u,v;A(\xi)]_{\text{hyper. ener.}} & = 
\int_{-\infty}^{\infty}\bigg(  
\Big( 
(\alpha\beta - 1) ( \lambda\gamma\exp(\delta u) + \delta\kappa\exp(\gamma v) )
+ \tfrac{1}{2} \delta \gamma \big( 
(\alpha-1) (u_t \pm\b u_x)^2 
\\&\qquad
+ (\beta-1) (v_t\pm\b v_x)^2
+ (u_t-v_t \pm\b(u_x-v_x))^2 \big)
\Big) A(x\pm \b t)
\\&\qquad
\pm 2\b \big( (\delta\beta -\gamma) v_t +(\gamma\alpha -\delta) u_t \big) A'(x\pm \b t)
+2\b^2\big((\delta -\gamma\alpha) u 
\\&\qquad
+ (\gamma -\delta\beta) v \big) A''(x\pm \b t) 
\bigg)\; dx
\end{aligned}
\end{equation}
which involves an arbitrary travelling wave function $A(\xi)$
and the light-cone derivatives $u_t\pm\b u_x$ and $v_t\pm\b v_x$
which represent the characteristic directions
for the semilinear system \eqref{uveqns} when $\a=\b$. 
This energy quantity is positive when $\alpha>1$, $\beta>1$, 
$\lambda\gamma>0$, $\delta\kappa>0$, $\delta\gamma>0$, 
and $A=\const>0$. 
Unlike parts (a) and (c), 
the appearance of an arbitrary function in the energy 
is not due to any linearity in the system \eqref{uveqns}
but instead comes from the exponential form of the nonlinear terms
$\f(v) = \kappa e^ {\mu v}$, 
$\g(u) = \lambda e^{\delta u}$, 
$\d(v) = \alpha \f(v)$, 
$\c(u) =  \beta \g(u)$. 
With this type of nonlinearity, 
the system \eqref{uveqns} is analogous to 
the Liouville wave equation $w_{tt}-\b^2 w_{xx} + \gamma \exp(w)=0$
for which it is known that all solutions $w(t,x)$ are functions only of 
the moving coordinate $x\pm\b t$. 
Consequently, 
an arbitrary function of $x\pm\b t$ is compatible with conservation of energy
in such systems. 

The conservation law \eqref{TXd1} in part (d) 
involves an arbitrary function of a linear combination of 
the light-cone derivatives $u_t\pm\b u_x$ and $v_t\pm\b v_x$
as given by $\zeta= \alpha u_t-v_t \pm\b(\alpha u_x-v_x)$.
Similarly to part (c),
the wave equations \eqref{ueqn} and \eqref{veqn} 
in the case $\c(u) =(1/\alpha)\g(u)$, $\d(v) =\alpha\f(v)$, $\a=\b$
can be combined to yield the ordinary linear wave equation 
\begin{equation}\label{linwaveeqnmassless}
(\alpha u -v)_{tt} - \b^2(\alpha u - v)_{xx} =0 . 
\end{equation}
This wave equation is well-known to admit any function of $\zeta$ 
as a conserved density. 
As a result, 
the corresponding conserved quantity for the semilinear system \eqref{uveqns} 
is given by 
\begin{equation}\label{lightlikeenermom}
\C[u,v;B(\zeta)]_{\text{ener./mom.}}  
= \int_{-\infty}^{\infty} B(\alpha u_t-v_t \pm\b(\alpha u_x-v_x))\; dx
\end{equation}
where $B$ is an arbitrary function of $\zeta$. 
When $B$ is linear in $\zeta$, 
this yields an elementary conserved quantity. 
When $B$ is nonlinear in $\zeta$, 
the conserved quantity is an energy-type quantity if $B$ is an even function
or a momentum-type quantity if $B$ is an odd function. 

Lastly, the Lagrangian case will be considered. 

Among the preceding conserved quantities, 
the ones that are admitted in the case of a Lagrangian formulation \eqref{uvLsys} consist of 
the energy \eqref{ener}, momentum \eqref{mom}, boost momentum \eqref{boostmom}, 
and hyperbolic energy \eqref{hyperener}. 
Their multipliers thereby correspond to variational symmetries
$\X=Q^u\parder{u}+Q^v\parder{v}$, 
which are respectively given by 
a time-translation \eqref{QIIIb3}, 
a space-translation \eqref{QIIIb2}, 
a Lorentz boost \eqref{QIIIb1}, 
and a generalized space-time translation combined with a generalized shift on $u,v$ \eqref{QIIIe1}. 

There are two additional conservation laws \eqref{TXf1} and \eqref{TXg1}, 
in part (f) and (g) respectively, 
holding in the massless Lagrangian case 
$\c=\d=0$, $\f(v)= \alpha v^{k-1}$, $\g(u)=\beta u^{-k-1}$, $\a=\b$ 
for which the nonlinearities are powers of $u$ and $v$. 
In this case, 
the wave equations \eqref{ueqn} and \eqref{veqn} are Euler-Lagrange equations
with respect to $v$ and $u$, 
where the Lagrangian is given by
\begin{equation}
L=-u_t v_t + \b^2 u_x v_x + \alpha\int v^{k-1} dv + \beta \int u^{-k-1} du
\end{equation}
with 
\begin{equation}
\int w^{n-1} dw 
= \begin{cases}
(1/n)w^{n} & n\neq 0\\
\ln(w) & n=0
\end{cases}.
\end{equation}
The multipliers \eqref{QIIIf1} and \eqref{QIb1} 
for these two conservation laws
correspond to variational symmetries 
$\X=Q^u\parder{u}+Q^v\parder{v}$ of the Lagrangian system. 
The first multiplier \eqref{QIIIf1} represents 
an infinitesimal rotation on $(u,v)$,
so the resulting conserved quantity 
\begin{equation}\label{rotcharge}
\C[u,v]_{\text{rot. charge}}  = \int_{-\infty}^{\infty} (u v_t - v u_t)\; dx
\end{equation}
is an $SO(2)$ rotation charge in the space $(u,v)$. 
The second multiplier \eqref{QIb1} represents 
an infinitesimal scaling on $(t,x,u,v)$,
and hence the resulting conserved quantity is a dilational energy-momentum
\begin{equation}\label{dilenermom}
\C[u,v]_{\text{ener.-mom.}}  = 
\int_{-\infty}^{\infty} \big( 
k t( v_t u_t +\b^2 v_x u_x ) + k x (v_x u_t + u_x v_t) +2(v u_t-u v_t)  +t(\alpha v^{k} -\beta u^{-k})
\big)\; dx . 
\end{equation}

\section{Concluding remarks}
\label{remarks}

The system of nonlinearly coupled wave equations \eqref{uveqns} 
has a fairly rich structure of low-order conserved quantities,
particularly when the wave speeds are equal, $\a=\b$. 
These quantities include energy, momentum, and boost momentum, 
plus a dilational energy-momentum and a $SO(2)$ rotation charge 
in the Lagrangian case 
(when the only nonlinearities are due to the coupling between $u$ and $v$). 
In two non-Lagrangian cases
(when the nonlinearities also comprise a self-coupling for both $u$ and $v$), 
additional energy-type conserved quantities arise, 
one involving arbitrary functions of 
the travelling wave coordinates $x\pm \b t$
and the other involving arbitrary functions of a linear combination of 
the light-cone derivatives $u_t\pm\b u_x$ and $v_t\pm\b v_x$. 

All of these conserved quantities will be useful in the analysis of 
solutions of this system. 
We will leave this analysis for elsewhere. 

An interesting problem for future work will be to look for 
higher-order conservation laws and higher-order symmetries,
as this will detect if the system \eqref{uveqns} is integrable 
for any special types of nonlinearities. 
In particular, 
this system is a natural two-component generalization of the Klein-Gordon equation 
which includes as special cases 
the Liouville equation, 
the sine-Gordon/sinh-Gordon equation, 
and the Tzetzeica equation,
each of which are integrable in the sense of possessing 
an infinite hierarchy of conservation laws and symmetries
(see \eg/ Ref.\cite{AblCla,AncBlu02a}).

\section*{Acknowledgements}

S.C. Anco is supported by an NSERC research grant.

\end{document}